\documentclass[showpacs,preprint,aps]{revtex4}%

\usepackage{stmaryrd}
\usepackage{amstext}
\usepackage{amsfonts}
\usepackage{amsmath}
\usepackage{amssymb}
\usepackage{epsfig}
\usepackage{graphicx}%
\ifx\pdfoutput\relax\let\pdfoutput=\undefined\fi
\newcount\msipdfoutput
\ifx\pdfoutput\undefined\else
\ifcase\pdfoutput\else
\ifx\paperwidth\undefined\else
\ifdim\paperheight=0pt\relax\else\pdfpageheight\paperheight\fi
\ifdim\paperwidth=0pt\relax\else\pdfpagewidth\paperwidth\fi
\fi\fi\fi
\begin{document}

\title{Strengthen Weak Measurement with Conjugated Destructive Interference}

\author{Zi-Huai Zhang}
\affiliation{Key Laboratory of Quantum Information, University of
Science and Technology of China, CAS, Hefei, 230026, China}
\affiliation{Synergetic Innovation Center of Quantum Information and Quantum Physics, University of Science and Technology of China, Hefei, Anhui 230026, China}

\author{Geng Chen$\footnote{email:chengeng@ustc.edu.cn}$}
\affiliation{Key Laboratory of Quantum Information, University of
Science and Technology of China, CAS, Hefei, 230026, China}
\affiliation{Synergetic Innovation Center of Quantum Information and Quantum Physics, University of Science and Technology of China, Hefei, Anhui 230026, China}

\author{Xiao-Ye Xu}
\affiliation{Key Laboratory of Quantum Information, University of
Science and Technology of China, CAS, Hefei, 230026, China}
\affiliation{Synergetic Innovation Center of Quantum Information and Quantum Physics, University of Science and Technology of China, Hefei, Anhui 230026, China}

\author{Jian-Shun Tang}
\affiliation{Key Laboratory of Quantum Information, University of
Science and Technology of China, CAS, Hefei, 230026, China}
\affiliation{Synergetic Innovation Center of Quantum Information and Quantum Physics, University of Science and Technology of China, Hefei, Anhui 230026, China}

\author{Wen-Hao Zhang}
\affiliation{Key Laboratory of Quantum Information, University of
Science and Technology of China, CAS, Hefei, 230026, China}
\affiliation{Synergetic Innovation Center of Quantum Information and Quantum Physics, University of Science and Technology of China, Hefei, Anhui 230026, China}

\author{Yong-Jian Han}
\affiliation{Key Laboratory of Quantum Information, University of
Science and Technology of China, CAS, Hefei, 230026, China}
\affiliation{Synergetic Innovation Center of Quantum Information and Quantum Physics, University of Science and Technology of China, Hefei, Anhui 230026, China}

\author{Chuan-Feng Li$\footnote{email:cfli@ustc.edu.cn}$}
\affiliation{Key Laboratory of Quantum Information, University of
Science and Technology of China, CAS, Hefei, 230026, China}
\affiliation{Synergetic Innovation Center of Quantum Information and Quantum Physics, University of Science and Technology of China, Hefei, Anhui 230026, China}

\author{Guang-Can Guo}
\affiliation{Key Laboratory of Quantum Information, University of
Science and Technology of China, CAS, Hefei, 230026, China}
\affiliation{Synergetic Innovation Center of Quantum Information and Quantum Physics, University of Science and Technology of China, Hefei, Anhui 230026, China}

\begin{abstract}
Standard weak measurement (SWM) has been proved to be a useful ingredient for measuring small longitudinal phase shifts. [$\emph{Phys. Rev. Lett.}$ $\textbf{111}$, 033604 (2013)]. In this letter, we show that with specific pre-coupling and postselection, destructive interference can be observed for the two conjugated variables, i.e. time and frequency, of the meter state. Using a broad band source, this conjugated destructive interference (CDI) can be observed in a regime approximately 1 attosecond, while the related spectral shift reaches hundreds of THz. This extreme sensitivity can be used to detect tiny longitudinal phase perturbation. Combined with a frequency-domain analysis, conjugated destructive interference weak measurement (CDIWM) is proved to outperform SWM by two orders of magnitude.
\end{abstract}


\maketitle

\section{Introduction}

High sensitivity detection of longitudinal phase shift is essential in metrology, such as accurate
distance measurements, timing synchronization and detection of gravitational waves \cite{Kim,Lee,Abbott}.
The standard tool is an interferometer
with a balanced homodyne detection \cite{Caves}. It requires
a coherent source and the precision is dominated by the
intrinsic quantum noise \cite{Yurke}. Recently, weak measurement has attracted
extensive attention and academic interest due to
its practical and potential applications in observing very small physical
effects \cite{Aharonov1,Aharonov2}. When weak measurements are judiciously combined with
preselection and postselection, they lead to a weak-value-amplification
phenomenon and give access to an experimental
sensitivity beyond the detector's resolution \cite{Brunner1,Brunner2,Hosten,Dixon,Starling1}. Brunner $\emph{et al.}$ compared the standard interferometer proposal with a weak measurement proposal. Their results evidently showed that when the purely imaginary weak value was exploited, weak measurement surpassed the standard interferometer proposal by several orders of magnitude \cite{Brunner3}. With this method, Xu $\emph{et al.}$ experimentally realized a very precise phase estimation utilizing
white light from a commercial light-emitting diode
(LED) \cite{Xu,Li}.

In Xu's scheme, the working regime is selected to make the coupling strength approximately 0, where postselection only slightly reshapes the wave-package pattern to obtain a larger shift considering the mean value of the spectrum. Here, we propose an improved scheme to measure longitudinal phase shifts. With a fixed pre-coupling of the system and the meter, the joint state becomes entangled. Afterwards with a specific postselection state, destructive interference can be observed in both the time and frequency domains using a broad light source. The phase shift is much smaller than the wave length and it still gets a destructive interference. In this process approximately 1 attosecond, the spectrum shifts a considerable amount decided by the postselection state. Prospectively this effect can be used to perform an ultra-sensitive measurement on tiny longitudinal phase perturbation.
%Combined with a frequency-domain analysis, CDIWM can achieve a resolution outperforming SWM proposal by two orders of magnitude.

\section{Generation of CDI}
We consider a measurement scenario involving a physical quantum state consisting of a system state $\lvert \psi \rangle$ and a meter state $\lvert g(x) \rangle$ where $x$ represents the space coordinate. The Hamiltonian of the system-meter combination is $H=kAP$. $A$ is an obeservable of the system and $P$ is the momentum of the meter. $A$ has two eigenstates $\lvert 0 \rangle$ and $\lvert 1 \rangle$ which satisfy $A\lvert 0 \rangle = \lvert 0 \rangle$ and $A\lvert 1 \rangle = -\lvert 1 \rangle$.

The system-meter evolution due to the interaction $H$ can be described by a unitary operator $U=e^{-iH\Delta t}$ with $\Delta t $ being the duration of the interaction. If the initial system state $\lvert \psi \rangle$ is a superposition of $\lvert 0 \rangle$ and $\lvert 1\rangle$,  i.e.  $\lvert \psi \rangle= \alpha \lvert 0 \rangle+ \beta \lvert 1\rangle$, the system-meter combination after the interaction is
\begin{equation}
\lvert \Psi \rangle= U\lvert \psi \rangle \lvert g(x)\rangle = \alpha  \lvert g(x+\tau)\rangle \lvert 0 \rangle +\beta  \lvert g(x-\tau)\rangle \lvert 1 \rangle,
\end{equation}
where $\tau=k\Delta t$. In this letter, we are interested in measuring the ultrasmall $\tau$ precisely.

For clarity, we consider the quantum state to be a Gaussian wave-package. The polarization of the photons serves as the system and the time or the frequency freedom serves as the meter. The interaction $H$ is introduced by a birefringent crystal which has different refractive indexes for the horizontal polarization $\lvert H \rangle$ and the vertical polarization $\lvert V \rangle$. In the scheme by Brunner $\emph{et al.}$, they first derive the wave-package in the time domain and then Fourier transform it into the frequency domain. Here, we will analyze the quantum system in both the time and frequency domains.

In the frequency domain, the wave function of the meter is represented by  $f(\omega)=(\pi \delta^2)^{-1/4}exp[-(\omega-\omega_0)^2/2\delta^2]$. Two nearly orthogonal circular polarization states $\lvert  \psi \rangle=\frac{1}{\sqrt{2}}(\lvert H \rangle+i\lvert V \rangle)$ and $\lvert  \phi \rangle=\frac{1}{\sqrt{2}}(ie^{i\epsilon}\lvert H \rangle+e^{-i\epsilon}\lvert V \rangle)$ are used for the preselection and the postselection processes. Initially, the system-meter combination is a separable state
\begin{equation}
\lvert \Phi \rangle = \int d\omega \frac{1}{\sqrt{2}}f(\omega)[\lvert H \rangle+i\lvert V\rangle]\lvert \omega \rangle].
\end{equation}

After the pre-coupling by inducing an initial delay $\tau$, the polarization and the frequency freedom are entangled
\begin{equation}
\label{initial}
\lvert  \Psi \rangle=\int d\omega \frac{1}{\sqrt{2}}f(\omega)[e^{i\omega\tau}\lvert H \rangle + ie^{-i\omega\tau}\lvert V\rangle]\lvert \omega \rangle.
\end{equation}

With a postselection of polarization, the meter state becomes (after normalization)
\begin{equation}
\lvert T \rangle=\frac{1}{\sqrt{P}}\int d\omega \frac{i}{2}f(\omega)[e^{i(\omega \tau -\epsilon)}-e^{-i(\omega \tau -\epsilon)}]\lvert \omega \rangle,
\end{equation}
with the postselection probability
\begin{equation}
\label{prob}
P=0.5\{1-exp(-\delta^2 \tau^2)cos[2(\omega_0 \tau-\epsilon)]\}.
\end{equation}

The postselected wave-package in frequency domain is $f(\omega)sin(\omega \tau -\epsilon)$ (without normalization), so the frequency propability distribution is
\begin{equation}
\label{spectrum}
S(\omega)=sin^2(\omega \tau -\epsilon) \lvert f(\omega)  \rvert^2.
\end{equation}

The spectral shift $\Delta \omega$  is calculated as the shift of the mean spectrum
\begin{equation}
\label{spectrumshift}
\Delta \omega= \frac{\int S(\omega)\omega d\omega}{\int S(\omega) d \omega}-\omega_0 =\frac{\tau \delta^2}{2P}exp(-\delta^2 \tau^2)sin[2(\omega_0 \tau-\epsilon)].
\end{equation}

The temporal probability distribution is $T(t)=\lvert  F[f(\omega)e^{i(\omega \tau -\epsilon)}-f(\omega)e^{-i(\omega \tau -\epsilon)}] \rvert^2$ where $F[ *]$ denotes the Fourier transform.

The SWM scheme has to satisfy $ \lvert \omega \tau/ \epsilon \rvert \ll 1$ in order to keep the interaction weak. Eq. (\ref{spectrum}) is derived to be
\begin{equation}
\label{SWMspectrum}
S(\omega)=\epsilon^{2} \lvert f(\omega)  \rvert^2.
\end{equation}

\begin{figure}[htbp]
\centering
\includegraphics[width=6in]{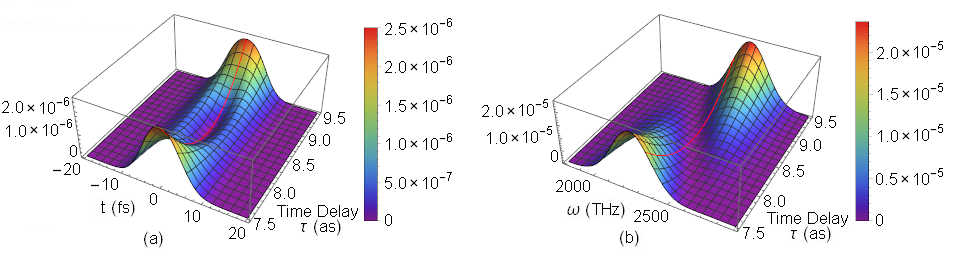}
\caption{The wave package in both (a) time and (b) frequency domains after postselection in the CDI regime. The pre- and post-selected states are $\lvert \psi \rangle=\frac{1}{\sqrt{2}}(\lvert H \rangle+i\lvert V \rangle)$ and $\lvert \phi \rangle=\frac{1}{\sqrt{2}}(ie^{0.02i}\lvert H \rangle+e^{-0.02i}\lvert V \rangle)$. At the central point of 8.5 as, the central components of both time and frequency domains are extinct due to destructive interference. The red lines in (a) and (b) indicate the central time and spectrum points, respectively.}
\label{Interaction}
\end{figure}

Within the weak interaction range where $\tau$$\rightarrow 0$, there is no destructive interference from Eq. (\ref{SWMspectrum}) so the spectral shift can be calculated from the purely imaginary weak value. However, from Eq. (\ref{spectrum}), when $\epsilon$$>$0 there is always a small $\tau_{s}$ satisfying $\omega_{0} \tau_{s}-\epsilon  \approx 0$ so that destructive interference can be observed around $\tau_{s}$. The wave-package evolution of the time and frequency domains around $\tau_{s}$ are shown in Fig. \ref{Interaction}. The initial meter state is normally prepared in a Gaussian superposition with the mean frequency $\omega_0$ to be 2350 THz and the width $\delta$ to be 200 THz, hence $\tau_{s}$ is calculated as 8.5 as. As shown in Fig. \ref{Interaction}(a), the central part of the time-domain wave package destructs to display a bimodal pattern at first and then recovers to a Gaussian distribution. While for frequency-domain in Fig. \ref{Interaction}(b), as the extinction point sweeps from the high to the low frequency with increasing $\tau$, the dominant wave package shifts from the low to high frequency range. In the center point around 8.5 as, the spectrum splits into two peaks. As the low frequency peak falls, the high frequency peak rises and they are equivalent at 8.5 as.

\section{Comparison of two Schemes}
From Fig. \ref{Interaction}, both the time and the frequency domains destructive interference can be observed in the chosen regime. Considering the current ultimate temporal-resolution which is on the order of several picoseconds, it is difficult to observe the time-domain destructive interference experimentally. In the following analysis we will mainly focus on the measurement in frequency-domain. The peak-position of the start and end time points can be distinguished intuitively in Fig. \ref{Interaction}(b), indicating a considerable spectral shift during 7.5 to 9.5 as. This means that making use of the CDI effect should give rise to a better resolution when measuring a tiny longitudinal phase change.
 
\begin{figure}[htbp]
\centering
\includegraphics[width=6in]{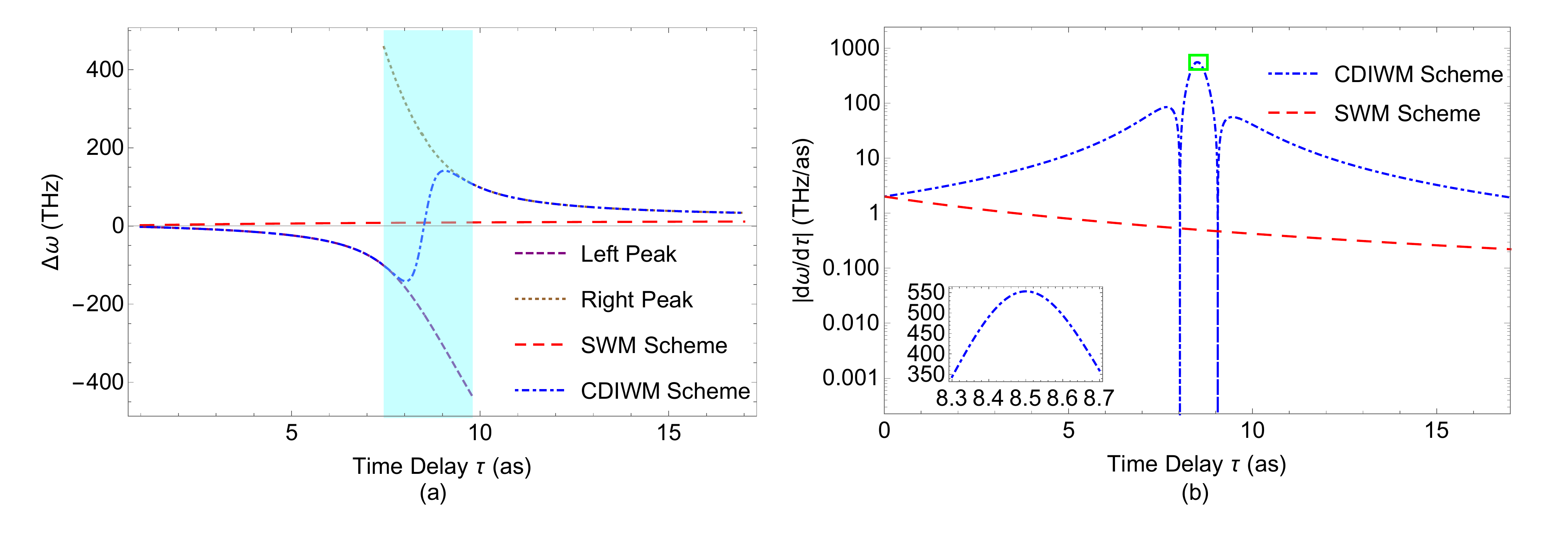}
\caption{(a) Spectral shifts in the CDIWM and the SWM schemes. The dot-dashed blue and long-dashed red lines are the mean spectral shifts in the CDIWM scheme and the SWM scheme respectively. The short-dashed violet and dotted brown lines represent the peak position shifts of the two parts divided by the extinction point in the CDIWM scheme. (b) Spectral shift rates in the CDIWM and the SWM schemes, calculated as the slope of corresponding spectral shift in (a). The dot-dashed blue line represents CDIWM and the long-dashed red line represents SWM. The postselection angle $\epsilon$ is set to be $0.02$ in CDIWM and $-0.02$ in SWM. The mean frequency $\omega_0$ is 2350 THz and the width $\delta$ is 200 THz. The green box and the inset identify the working range of CDIWM.}
\label{SpectrumSlope}
\end{figure}

 It has been proved that the longitudinal phase change can be precisely detected by measuring the spectral shift \cite{Brunner3}. In Fig. \ref{SpectrumSlope}, we plot the spectral shifts as well as the shift rates of the two schemes according to Eq. (\ref{spectrumshift}). For the SWM scheme, the parameters are set to be identical to Xu's scheme and the results are consistent with his work. Fig. \ref{SpectrumSlope}(a) shows the spectral shifts of both the CDIWM scheme and the SWM scheme varying $\tau$. In the CDIWM scheme, the frequency domain wave-package splits into two peaks in the light blue area, either of which the peak position is also shown. It can be seen that the SWM scheme has only a spectral shift of several THz while the CDIWM scheme can reach a shift of several hundreds THz. Even more noteworthy is that there is a very steep spectral shift around 8.5 as, which means an extremely high sensitivity when measuring the phase perturbation.

Particularly, when there is a perturbation on the relative phase, we require a change on the meter as large as possible. This can be characterized by the spectral shift rate with respect to the phase perturbation or the equivalent time delay. Fig. \ref{SpectrumSlope}(b) shows that the spectral shift rate of the CDIWM scheme is far larger than that of the SWM scheme. Working on the most sensitive point, the CDIWM scheme can reach a shift rate beyond the SWM scheme by two orders of magnitude. To obtain the best sensitivity and a stable interference pattern with the CDIWM scheme, the working point should be accurately stabilized in a small time regime (the green box and the inset in Fig. \ref{SpectrumSlope}(b)). Existing techniques can provide the required accuracy to set the initial work point \cite{Abbott,Luis}.

\begin{figure}[htbp]
\centering
\includegraphics[width=3in]{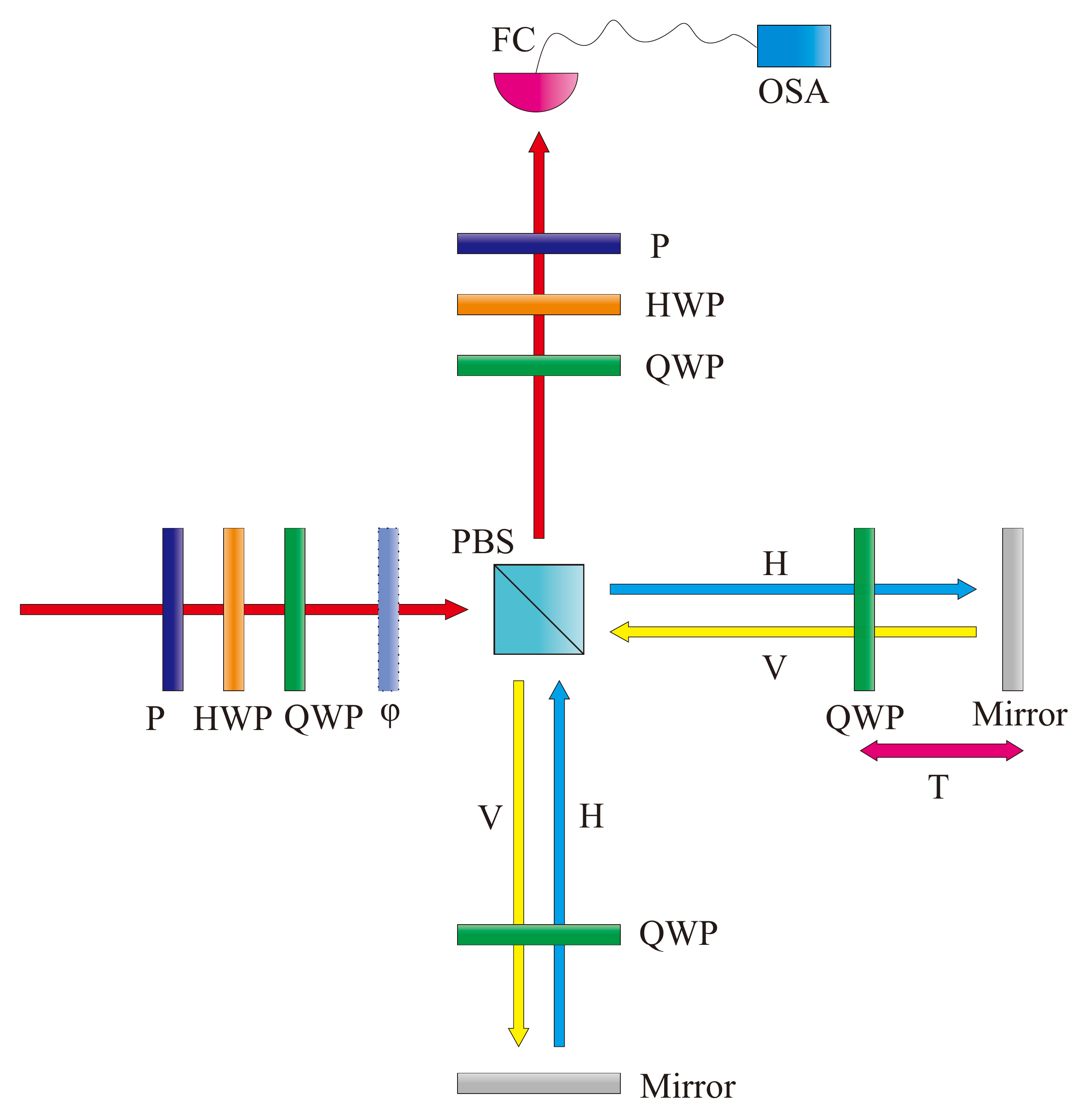}
\caption{The proposed experimental setup for the CDIWM scheme. The sets of polarizers (P), half wave plates (HWP) and quarter wave plates (QWP) are used for the pre- and postselection procedures. A PBS splits the input beam into two orthogonal linear polarization beams propagating along different arms of the interferometer. The QWP together with the mirror rotate the polarization by 90$^\circ$ so as to recombine the two beams at the same PBS. The moveable mirror is used to tune the initial coupling. The fiber collector (FC) collects the output beam whose spectrum is measured using an optical spectrum analyzer (OSA). $\phi$ denotes the tiny longitudinal phase perturbation between different polarization components.}
\label{setup}
\end{figure}

Fig.  \ref{setup} shows the proposed experimental setup based on a Michelson Interferometer (MI). The beam splitter (BS) in the traditional MI is substituted by a polarizing beam splitter (PBS). The specific pre-coupling required by CDIWM is realized by adjusting the length difference of the two arms as well as analyzing the postselected spectrum. Initially, the two arms are set to be equal. As the mirror moves, the postselected spectrum shifts according to Fig. \ref{Interaction}(b) . The most sensitive point is confirmed when the extinct point is in the middle of the spectrum. Afterwards all the optical elements are locked so that the system is on standby for the measurement.

%The sets of polarizers (P), half wave plates (HWP) and quarter wave plates (QWP) are used for the pre- and postselection procedures. Each of the two arms contains a QWP and a mirror. The QWP together with the mirror rotate the polarization by 90$^\circ$. The position of the mirror is adjustable so the relative phase delay can be tuned easily. FC, OSA and PBS denote fiber collector, optical spectrum analyzer and polarizing beam splitter, respectively. The phase $\phi$ denotes the tiny longitudinal phase perturbation between different polarization components to be measured.

%Tuning the interferometer to the working point of CDIWM can be realized by analyzing the postselected spectrum. The two arms are set to be equal initially. Then change the position of one mirror and measure the postselected spectrum.  A varying spectrum same as Fig. \ref{Interaction}(b) is expected. The most sensitive point is confirmed when the extinct point is in the middle of the spectrum. Afterwards all the optical elements are locked so that the system is on standby for the measurement.

The ultimate detectable longitudinal phase perturbation is estimated below. In the SWM scheme, when $ \lvert \omega \tau/ \epsilon \rvert \ll 1$ the spectral shift can be readily obtained from weak value defined as $A_w=\langle \phi \lvert A \rvert \psi \rangle/ \langle \phi \lvert \psi \rangle$. The ordinary weak value is calculated to be $icot(\epsilon)$ so the spectral shift is
\begin{equation}
\label{metershift}
\Delta \omega \rightarrow \tau Im[A_w]\delta^2=cot(\epsilon) \delta^2 \Delta\tau \approx \frac{\delta^2 \Delta \tau}{\epsilon}
\end{equation}
with a longitudinal phase perturbation $\Delta\tau$. The ultimate resolution limit of the SWM scheme is given by \cite{Brunner3}
\begin{equation}
\label{SWMresolution}
\tau>\frac{\lvert \epsilon \rvert \Delta \Omega}{\delta^2},
\end{equation}
where $\Delta \Omega$ is the resolution of the OSA. (There is a factor of 2 difference compared to the scheme by Brunner $\emph{et al.}$ due to a slight different in the representation of meter.) In the CDIWM scheme, the working regime is restricted to the CDI regime so the spectral shift can be calculated as
\begin{equation}
\Delta \omega \rightarrow \frac{d\Delta \omega}{d\tau}\lvert_{\tau \rightarrow \frac{\epsilon}{\omega_0}}\Delta\tau  \approx 2\frac{\omega_0^2}{\epsilon}\Delta\tau.
\end{equation}

The resolution for the CDIWM scheme is similarly analyzed to be
\begin{equation}
\label{CDIWMresolution}
\tau>\frac{\lvert \epsilon \rvert \Delta \Omega}{2\omega_0^2}.
\end{equation}

\begin{figure}[htbp]
\centering
\includegraphics[width=6in]{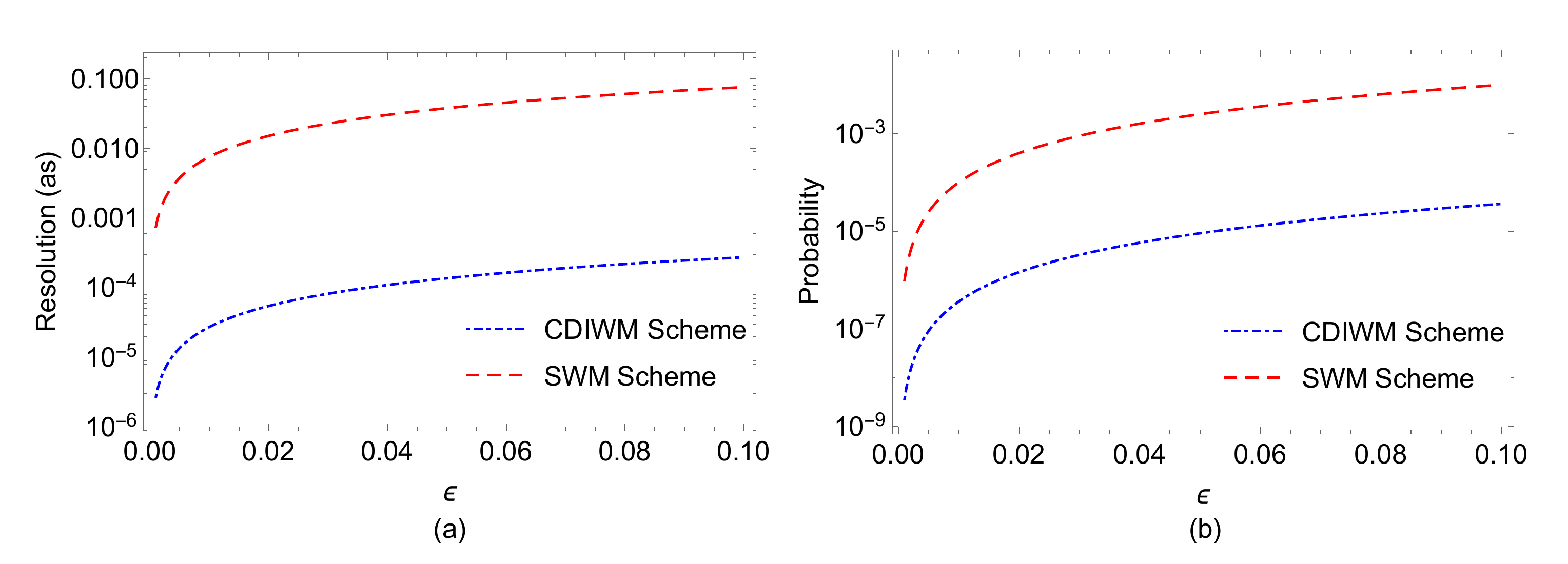}
\caption{The comparison of CDIWM and SWM with respect to the postselection angle $\epsilon$. (a) The resolution of two schemes. (b) Postselection probabilities of two schemes. The dot-dashed blue and long-dashed red lines represent the CDIWM scheme and the SWM scheme, respectively. }
\label{ResolutionSelection}
\end{figure}

From the above analysis, CDIWM can achieve a significantly higher resolution. With a visible broad band source, $\omega_0$ is on the order of several thousands THz and $\delta$ is on the order of several hundreds THz so the resolution can be improved by two orders of magnitude. For example, we consider a light beam centering at $\omega_0=2350$ THz ($\lambda \approx 800$ nm) with $\delta =200$ THz ($\delta \lambda \approx 100$ nm) and a OSA having a spectral resolution of 0.01 nm at $\lambda=800$ nm. The achievable resolution is shown in Fig. \ref{ResolutionSelection}(a) with varying $\epsilon$ calculated from Eq. (\ref{SWMresolution}) and Eq. (\ref{CDIWMresolution}). When $ \epsilon  \rightarrow0$, CDIWM has the ability to detect a time-domain perturbation on the order of $10^{-5}$ as.

However, this improvement is at a cost that much more photons are discarded in the postselection process. In the SWM scheme, the postselection probability $P$ is approximately the probability of simply postselecting $\lvert \psi \rangle$ by $\lvert \phi \rangle$ which leads to $P_{SWM} \approx \epsilon^2$. In the CDIWM scheme, we require $\omega_0 \tau -\epsilon \approx 0$ and the working regime to be $\tau \approx \epsilon/\omega_0$ so the postselection probability now becomes $P_{CDIWM}\approx \delta^2 \epsilon^2/(2\omega_0^2)$. Fig. \ref{ResolutionSelection}(b) shows the postselection probabilities of the two schemes with respect to different $\epsilon$ calculated from Eq. (\ref{prob}).

To sum up, by using the CDIWM scheme the resolution is improved by two orders of magnitude with the same orders of decreasing in the postselection probability. Specifically, we can achieve a factor of $2\omega_0^2/\delta^2$ amplification for resolution while decreasing the probability by a factor of $\delta^2/(2\omega_0^2)$ at the same time using the CDIWM scheme instead of the SWM scheme. However, considering the SWM scheme exclusively, an improvement by two orders of magnitude leads to a decreasing of postselection probability by four rather than two orders of magnitude.

\section{Discussion}
\label{Discuss}
By pre-coupling the system and meter and postselecting with specific state, CDI occurs and leads to an improved measurement sensitivity accompanying by a lower postselection probability. This amplification and rejection processes are very similar to weak-value-amplification effect, however, the physical description of CDIWM largely differs from SWM.

\begin{figure}[htbp]
\centering
\includegraphics[width=5in]{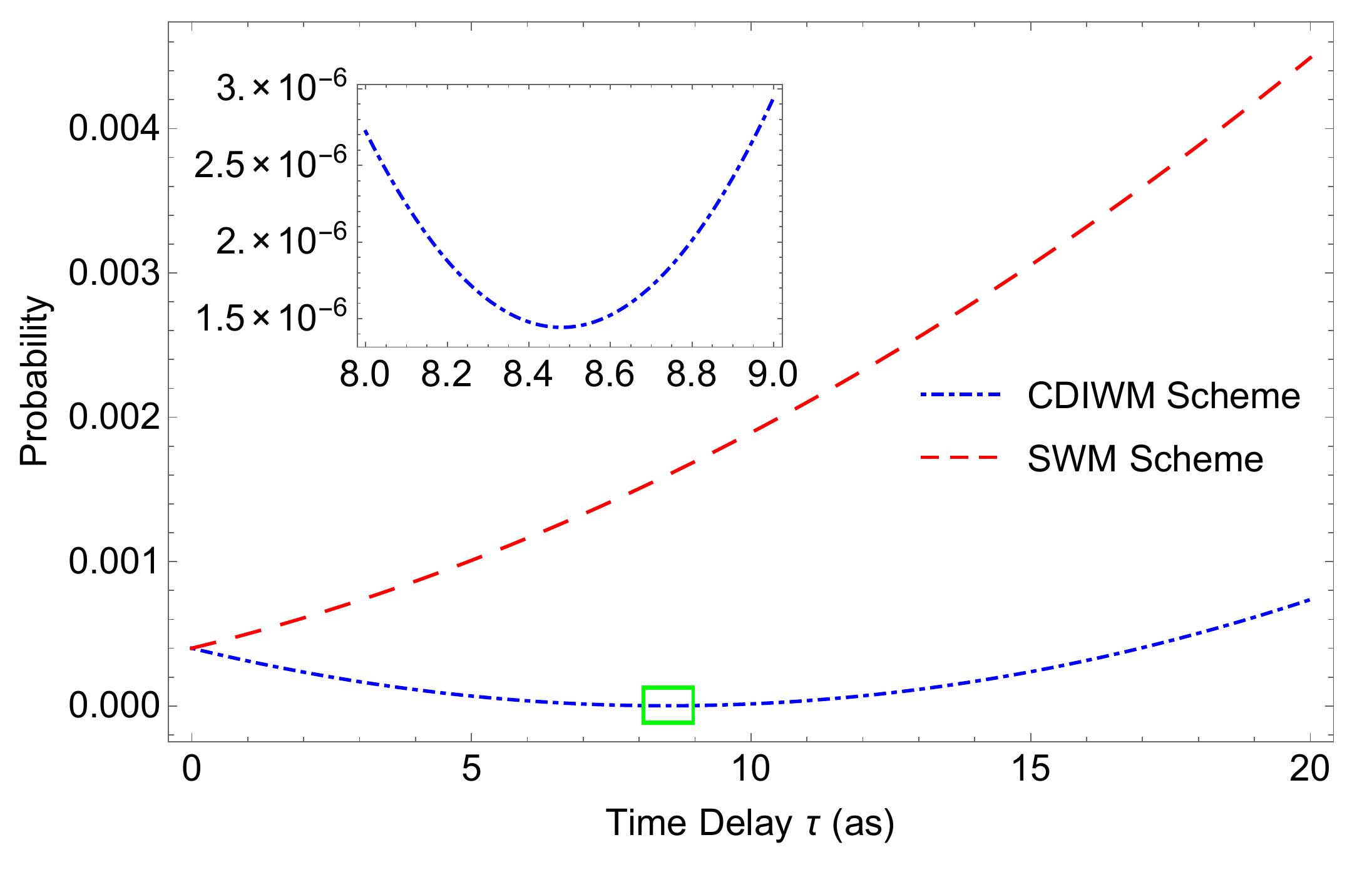}
\caption{The postselection probability curves of SWM and CDIWM with the same parameters in Fig. \ref{SpectrumSlope}. The dot-dashed blue line represents CDIWM and the long-dashed red line represents SWM. The green box and the inset indentify the working range of CDIWM. }
\label{ProbComp}
\end{figure}

In SWM, within the so called weak interaction regime \cite{Boyd}, the probability correction due to the interaction has a linear relationship with the first order weak value. The shift of meter can be calculated according to Eq. (\ref{metershift}). However, this is not the case in CDIWM. Fig. \ref{ProbComp} shows the postselection probability curves of SWM and CDIWM calculated from Eq. (\ref{prob}). The pre-coupling process shifts the working regime to $\tau_{s} = \epsilon/\omega_0$ which is out of the weak interaction regime and reaches the local minimum of postselection probability (the green box and the inset in Fig. \ref{ProbComp}).

On the other hand, this pre-coupling process makes an entangled initial state of the system and the meter according to Eq. (\ref{initial}). As a result, the weak value can not be well defined here and the weak interaction approximation $\langle \phi \lvert e^{-iAP\tau}\rvert\psi \rangle \approx \langle \phi \lvert \psi \rangle e^{-iA_w P \tau}$ doesn't hold anymore in the working regime of CDIWM, where $A_{w} \omega \tau \approx 1$ . 

Weak measurement can achieve a better resolution while discarding a large proportion of resources. Considering both the amplification effect and the loss due to postselection, the ultimate precision or signal-to-noise ratio can not outperform the classical metrology \cite{Combes,Gauger}. This issue also occurs in the CDIWM scheme because the existence of CDI leads to a further lower postselection. However, some recent proposals have proved that the power recycle technique can be used in weak measurement, thus recovering the inefficiency due to low postselection \cite{Kwiat1, Kwiat2}. Adding a partially transmitting mirror to make the interferometer a resonant optical cavity, all the input photons will exit the interferometer with the amplified signal.

\section{Conclusion}
We propose an improved weak measurement scheme to detect tiny perturbation to longitudinal phase in this letter. By pre-coupling the system and meter state and selecting a specific postselection state, a novel CDI effect can be observed when a broad band light source is used. From the calculation we find that when CDI occurs, the sensitivity of the spectral shift to longitudinal phase perturbation significantly increases. Respect to the same tiny perturbation on the longitudinal phase, the spectral shift can be amplified by a factor of several hundreds compared to the SWM scheme. Our results also outperform coherent light phase weak measurements \cite{Starling2} and currently are significantly better than quantum metrology technology measurements using N00N and squeezed states \cite{Krischek}, which are still in the process of solving experimental problems \cite{Thomas}. Taking advantages of these characteristics, when a physical effect is coupled to a longitudinal phase, it can be precisely estimated through this CDIWM scheme.
%\clearpage

{\bf  Acknowledgments}

 This work was supported by the National Basic Research Program of China (Grants No. 2013CB933304), the Strategic Priority Research Program (B) of the Chinese Academy of Sciences (Grant No. XDB01030300), National Natural Science Foundation of China (Grant Nos. 91536219, 61308010), the Fundamental
Research Funds for the Central Universities (Grant No. WK2030020019, WK2470000011).


\begin{thebibliography}{xx}
\bibitem{Kim} J. Kim, J. A. Cox, J. Chen, and F. X. Kartner, Drift-free femtosecond timing synchronization of remote optical and microwave sources. Nat. Photon. \textbf{2}, 733 (2008).
\bibitem{Lee} J. Lee, Y.-J. Kim, K. Lee, S. Lee, and S.-W. Kim, Time-of-flight measurement with femtosecond light pulses. Nat. Photon.
\textbf{4}, 716 (2010).
\bibitem{Abbott} B. P. Abbott \emph{et al.} Observation of gravitational waves from a binary black hole merger. Phys. Rev. Lett. \textbf{116}, 061102 (2016).
\bibitem{Caves} C. M. Caves, Quantum-mechanical noise in an interferometer. Phys. Rev. D 23, 1693 (1981).
\bibitem{Yurke} B. Yurke, S. L. McCall, and J. R. Klauder, SU(2) and SU(1,1) interferometers. Phys. Rev. A 33,
4033 (1986).
\bibitem{Aharonov1} Y. Aharonov, D. Z. Albert, and L. Vaidman, How the result of a measurement of a component of the spin of a spin-1/2 particle can turn out to be 100. Phys. Rev. Lett. \textbf{60}, 1351 (1988).
\bibitem{Aharonov2} Y. Aharonov and L. Vaidman, in Time in Quantum
Mechanics, edited by R. S. M. J. G. Muga and I. Egusquiza (Springer-Verlag, Berlin, 2002).
\bibitem{Brunner1} N. Brunner, V. Scarani, M. Wegmuller, M. Legr$\acute{e}$, and N. Gisin, Direct measurement of superluminal group velocity and signal velocity in an optical fiber. Phys. Rev. Lett. \textbf{93}, 203902 (2004).
\bibitem{Brunner2} N. Brunner, A. Acin, D. Collins, N. Gisin, and V. Scarani, Optical telecom networks as weak quantum measurements with postselection. Phys. Rev. Lett. \textbf{91}, 180402 (2003).
\bibitem{Hosten} O. Hosten and P. Kwiat, Observation of the spin Hall effect of light via weak measurements. Science \textbf{319}, 787 (2008).
\bibitem{Dixon} P. B. Dixon, D. J. Starling, A. N. Jordan, and J. C. Howell, Ultrasensitive beam deflection measurement via interferometric weak value amplification. Phys. Rev. Lett. \textbf{102}, 173601 (2009).
\bibitem{Starling1} D. J. Starling, P. B. Dixon, A. N. Jordan, and J. C. Howell, Optimizing the signal-to-noise ratio of a beam-deflection measurement with interferometric weak values. Phys. Rev. A \textbf{80}, 041803 (2009).
\bibitem{Brunner3} N. Bruuner and C. Simon, Measuring small longitudinal phase shifts: weak measurements or standard interferometry? Phys. Rev. Lett. \textbf{105}, 010405 (2010).
\bibitem{Li} C. F. Li, X. Y. Xu, J. S. Tang, J. S. Xu, and G. C. Guo, Ultrasensitive phase estimation with white light. Phys. Rev. A \textbf{83}, 044102 (2011).
\bibitem{Xu} X. Y. Xu, Y. Kedem, K. Sun, L. Vaidman, C. F. Li, and Guang-Can Guo, Phase estimation with weak measurement using a white light source.
Phys. Rev. Lett. \textbf{111}, 033604 (2013).
\bibitem{Luis} L. J. Salazar-Serrano, A. Valencia, and J. P. Torres, Observation of spectral interference for any path difference in an interferometer. Opt. Lett. \textbf{39}, 4478 (2014).
\bibitem{Boyd} J. Dressel, M. Malik, F. M. Miatto, A. N. Jordan, and R. W. Boyd, Understanding quantum weak values: Basics and applications. Rev. Mod. Phys. \textbf{86}, 307 (2014).
\bibitem{Combes}C. Ferrie and J. Combes, Weak value amplification is suboptimal for estimation and detection. Phys. Rev. Lett. \textbf{112}, 040406 (2014).
\bibitem{Gauger}G. C. Knee and E. M. Gauger, When amplification with weak values fails to suppress technical noise. Phys.Rev.X \textbf{4}, 011032 (2014).
\bibitem{Kwiat1} J. Dressel , K. Lyons , A. N. Jordan, T. M. Graham, and P. G.
Kwiat, Strengthening weak-value ampliﬁcation with recycled photons. Phys. Rev. A \textbf{88}, 02382 1 (2013).
\bibitem{Kwiat2} K. Lyons, J. Dressel, A. N. Jordan, J. C. Howell, and P. G. Kwiat. Power-recycled weak-value- based Metrology. Phys. Rev. Lett. \textbf{114}, 170801 (2015).
\bibitem{Starling2} D. J. Starling, P. B. Dixon, N. S. Williams, A. N. Jordan,
and J. C. Howell, Continuous phase amplification with a Sagnac interferometer. Phys. Rev. A \textbf{82}, 011802(R) (2010).
\bibitem{Krischek} R. Krischek, C. Schwemmer, W. Wieczorek, H.
Weinfurter, P. Hyllus, L. Pezze and A. Smerzi, Useful multiparticle entanglement and sub-shot-noise sensitivity in experimental phase estimation. Phys.
Rev. Lett. \textbf{107}, 080504 (2011).
\bibitem{Thomas} N. Thomas-Peter, B. J. Smith, A. Datta, L. Zhang, U.
Dorner, and I. A. Walmsley, Real-world quantum sensors: evaluating resources for precision measurement. Phys. Rev. Lett. \textbf{107},
113603 (2011).

\end{thebibliography}
\end{document}